\documentclass{emulateapj}
\usepackage{graphicx}
\usepackage{amsmath}
\usepackage{natbib}

\shorttitle{Distribution of BH recoil velocities}
\shortauthors{Schnittman \& Buonanno}

\begin{document}

\title{The Distribution of Recoil Velocities from Merging Black Holes}
\author{Jeremy D.\ Schnittman and Alessandra Buonanno}
\affil{Department of Physics, University of Maryland,
College Park, MD 20742}

\begin{abstract}
We calculate the linear momentum flux from merging black holes (BHs) with
arbitrary masses and spin orientations, using the effective-one-body
(EOB) model. This model includes
an analytic description of the inspiral phase, a short merger, 
and a superposition of exponentially 
damped quasi-normal ringdown modes of a Kerr BH. By varying the
matching point between inspiral and ringdown, we can estimate the
systematic errors generated with this method. Within these
confidence limits, we find close agreement with previously reported
results from numerical relativity. Using a Monte Carlo implementation
of the EOB model, we are able to sample a large
volume of BH parameter space and estimate the distribution of
recoil velocities. For a range of mass ratios $1\le m_1/m_2 \le
10$, spin magnitudes of
$a_{1,2}=0.9$, and uniform random spin orientations, we find that a fraction
$f_{\rm 500}=0.12^{+0.06}_{-0.05}$ of binaries have recoil velocities
greater than 500 km/s and $f_{\rm 1000}=0.027^{+0.021}_{-0.014}$
have kicks greater than 1000 km/s. These velocities likely are capable of
ejecting the final BH from its host galaxy.
Limiting the sample to comparable-mass binaries with $m_1/m_2 \le 4$,
the typical kicks are even larger, with $f_{\rm 500}=0.31_{-0.12}^{+0.13}$ and
$f_{\rm 1000}=0.079^{+0.062}_{-0.042}$.
\end{abstract}


\maketitle

\section{INTRODUCTION}
\label{intro}

In the past year there has been remarkable progress made
in the field of numerical relativity (NR). One of the most exciting new
results is the calculation of the linear momentum flux generated by
the inspiral, merger, and ringdown of black hole (BH) binaries
\citep{herrm06,baker06,gonza06,herrm07,koppi07,campa07,gonza07,baker07}. 
Since the majority of this momentum flux is emitted during the merger
and ringdown, it is difficult to make definitive predictions
for the recoil using {\it only} analytic methods. However, in the non-spinning 
case, the post-Newtonian (PN) model~\citep{blanc05} 
has provided results consistent with NR all along the adiabatic 
inspiral; the effective-one-body (EOB) model 
can reproduce the total recoil, i.e., also the contribution 
from the ring-down phase, but with large
uncertainties~\citep{damour06}; perturbative 
models~\citep{sopue06} have also spanned the NR predictions. 
The recoil remains an ideal problem for which one can benefit 
enormously from accurate numerical simulations.

Recent NR results predicting very large recoil velocities $(\gtrsim 500
\mbox{ km/s})$ have also called attention to the potential
astrophysical importance of the recoil \citep{campa07,gonza07}. For
many models of dark matter
halo growth through hierarchical mergers, supermassive BHs at
the centers of such halos will inevitably also merge unless kicked
out of the gravitational potential well from a previous merger \citep{menou01}. The 
NR results have shown that this is indeed {\it possible}, 
but have not yet shown whether it is {\it probable} or not. 

As an initial investigation of this question, we have
applied the EOB approach \citep{BD1,BD2,DJS} 
to model the BH inspiral and ringdown phases including spin-orbit 
and spin-spin effects \citep{BCD}, described below in \S~2. Unlike
previous analytic approaches, we can now benefit from a rapidly
growing collection of numerical data, allowing us to calibrate our
model and quantify its uncertainties.
In \S~3, we compare a large number of these NR
simulations to our EOB predictions, agreeing within our conservative
error estimates. Having established a range of confidence in the EOB
model, we proceed in
\S~4 to show the results of Monte Carlo simulations
with a wide range of mass ratios, spin magnitudes,
and orientations. We thus provide the first estimates of the
distribution of recoil velocities from BH mergers,
and in \S~5 briefly discuss the astrophysical implications of these
results. 

\section{ANALYTIC MODEL OF BLACK HOLE  RECOIL}
\label{analytic_models}

For the two-body dynamics, we use the EOB model with spin-orbit 
and spin-spin terms included through 1.5PN and 2PN order, respectively, 
as described in \citet{BCD}. The non-spinning, conservative dynamics 
are computed through 3PN order \citep{DJS} and the radiation-reaction 
effects are included through 3.5PN order \citep{blanc04}. 
The initial conditions are taken to replicate an adiabatic, 
quasi-circular inspiral beginning at $r=10m$. 
Building on previous works \citep{BD2,BCD,damour06}, \citet{BCP}
were able to match the EOB inspiral-plunge waveform to a
linear combination of three Kerr quasi-normal ringdown (RD) modes,
obtaining qualitative agreement with the full NR inspiral-merger-RD
wave. In the model used here, we assume the final BH spin is given by
the linear scaling with mass ratio of \citet{gonza06}. 
After calculating the inspiral and RD dynamics, we determine the
linear momentum flux using the radiative multipole moments
described in \citet{thorn80}, including the leading-order radial
velocity and spin-orbit contributions to 
the individual modes \citep{kidde95}. These methods will be described
in greater detail in a companion paper.

The instantaneous transition from inspiral to ringdown turns out to be
rather sensitive to the point
of matching and it is only partially effective: the gravitational
wave (GW) frequency
resulting from the EOB matching grows too quickly during the
merger-RD transition, whereas the NR GW frequency increases more
gently. As a
way of estimating the optimal match point as well as the errors
associated with this ``prompt merger'' approximation, we calculate
the GW energy emitted in the inspiral relative to that of the
ringdown. Since the amplitudes of the individual multipole modes
typically increase rapidly with frequency during the inspiral, by matching to
the ringdown at a later time, the amplitudes of the excited RD modes and
thus the RD energy increases significantly. Similarly, if we match too early,
the resulting RD amplitudes are too small. Following \citet{flana98}, we scale
the RD energy according to $E_{\rm RD}(\mu) \propto \mu^2/m$ (we
define $m=m_1+m_2$, $\mu=m_1m_2/m$, and $\eta=\mu/m$, with $m_1>m_2$)
and scale the
inspiral energy to $E_{\rm ins}(\mu) \propto \mu \varepsilon$, where
$1-\varepsilon$ is the specific energy of a test particle
at the inner-most stable circular orbit (determined here via the
effective spin of the EOB model). Guided by NR results, we set $E_{\rm
ins}=E_{\rm RD}$ in the equal-mass case \citep{BCP}, and
use the scaling relations to determine the ``target'' inspiral
and RD fractions for other BH mass ratios and spins. 

We then vary the matching point in time, calculating the recoil for
each case, and requiring the fraction $E_{\rm ins}/(E_{\rm ins}+E_{\rm
  RD})$ to
agree with the target fraction within $\pm 15\%$. This typically
corresponds to a range of $r/m \approx 2.5-3.5$ and
$\omega_{\rm orb} \approx 0.1-0.2$ at the matching point. For our EOB
recoil predictions, we
take the mean kick velocity over the acceptable match range and define
1-$\sigma$ errors from the variance over this
range. By calibrating with 
the NR results and quantifying its uncertainties, the current 
EOB model can already be used to obtain interesting predictions 
of the recoil velocity.

\section{COMPARISON WITH NUMERICAL SIMULATIONS}
\label{compare_NR}

\begin{figure}
\begin{center} 
\scalebox{0.45}{\includegraphics{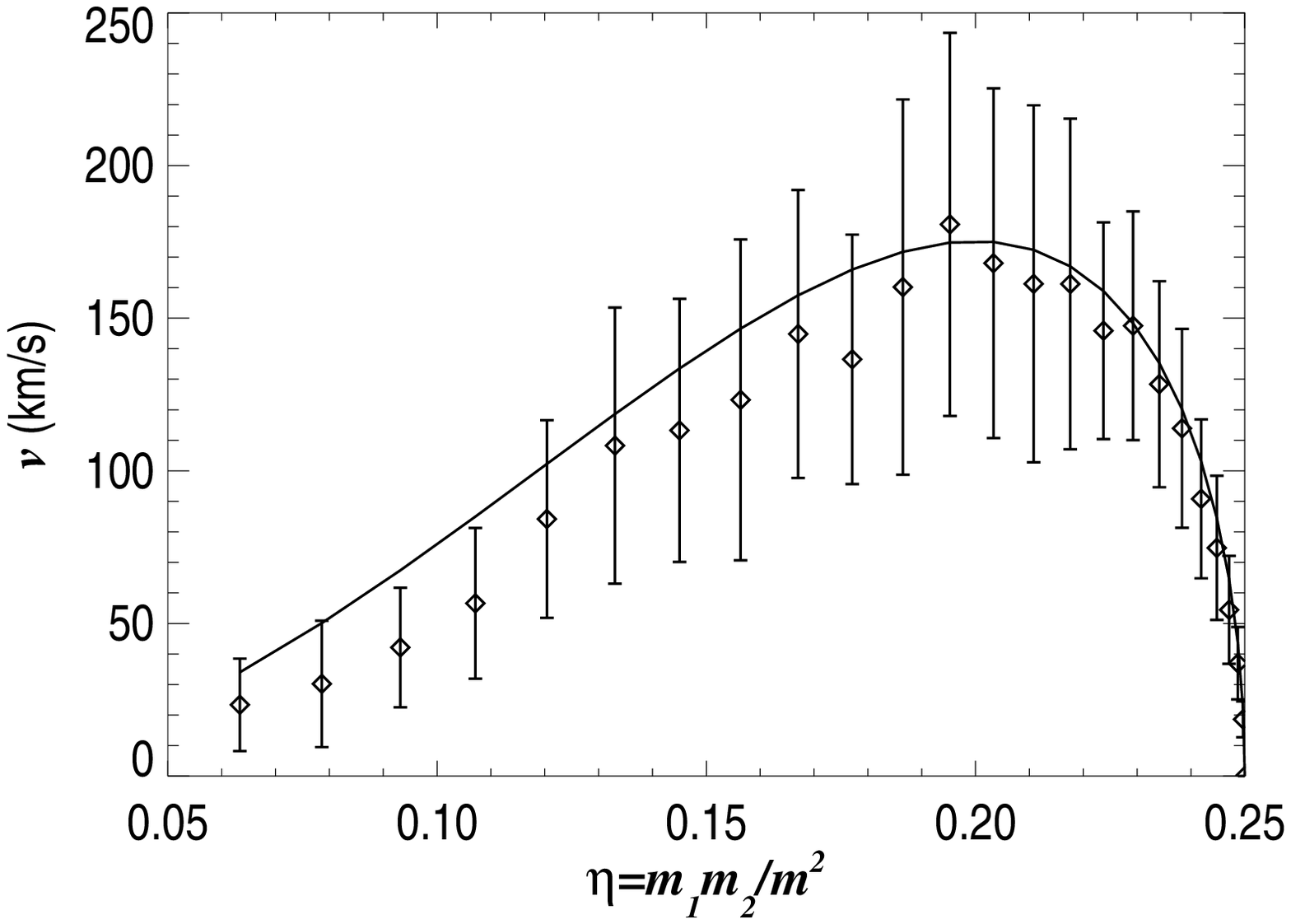}}\\
\scalebox{0.45}{\includegraphics{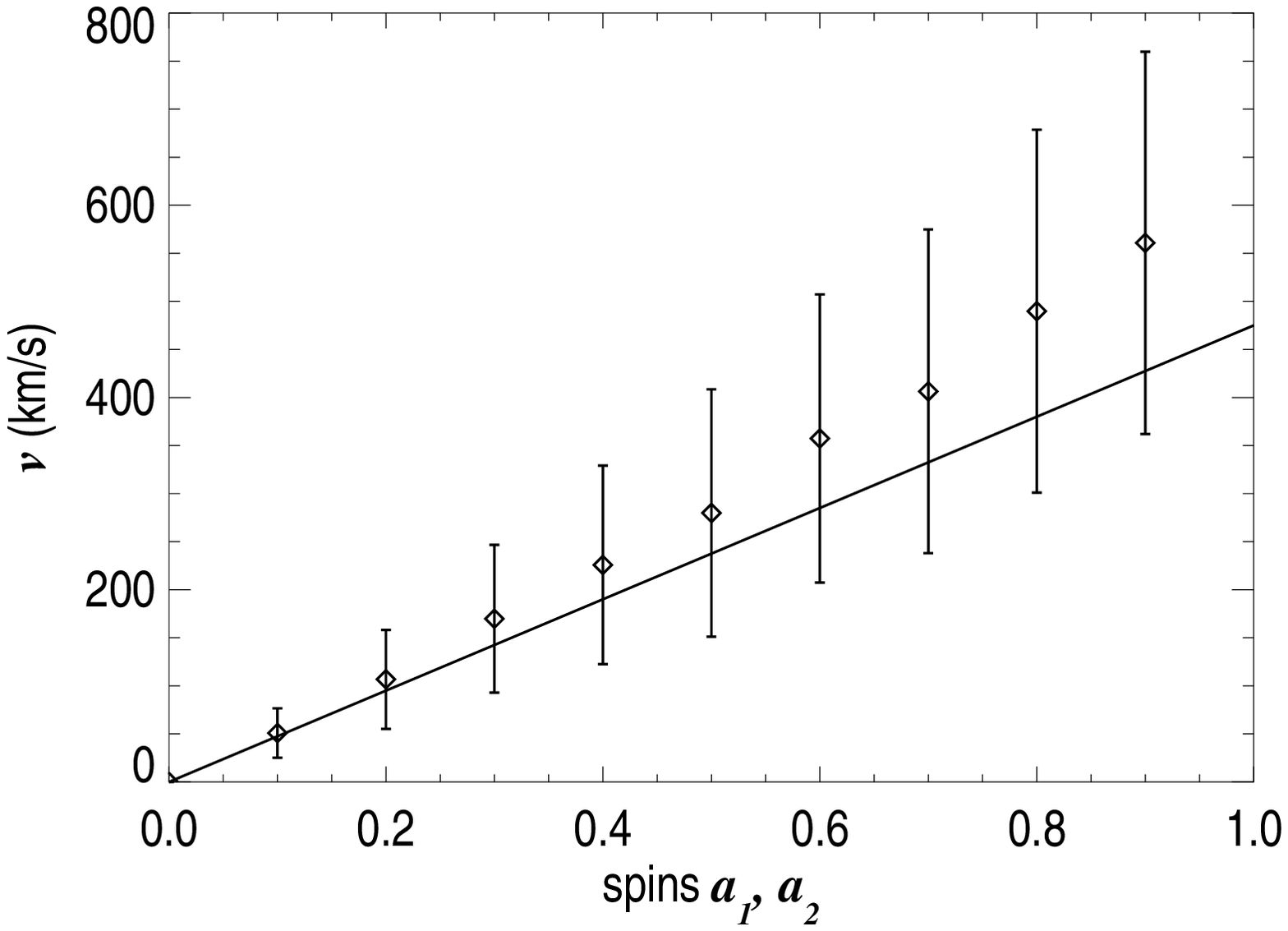}}
\caption{\label{plotone} \textit{Upper:} EOB predictions of the recoil
  velocity for non-spinning unequal-mass BHs, along with 1-$\sigma$
  errors. The solid line follows the
\citep{fitch83} scaling, normalized to the NR results in
\citet{gonza06}.
\textit{Lower:} Recoil for equal-mass binaries with spins aligned and
anti-aligned with the orbital angular momentum. The solid line is the
linear fit of \citet{herrm07}.}
\end{center}
\end{figure}

Unlike previous analytic calculations of the recoil,
we are now in the unique position of being able to compare our EOB
predictions directly with a growing collection of high-resolution NR
simulations and thereby establish a range of confidence in the results of our
analytic model. Beginning with the simplest case, in Figure
\ref{plotone} we show in the top panel the EOB recoil velocities for
non-spinning unequal-mass binaries (diamond
symbols with 1-$\sigma$ error bars), along with the classic formula of
\citet{fitch83}: $v(\eta) \propto \eta^2\sqrt{1-4\eta}$,
scaled to the results of \citet{gonza06} who find $v(\eta_{\rm
  max})=175$ km/s. In the bottom panel, we
show the results of equal-mass binaries with equal-magnitude spins
aligned and anti-aligned with the orbital angular momentum.
The solid
line is the linear scaling $v(a) = 475a$ km/s given by
\citet{herrm07}, where we define the dimensionless spin parameter
$a_{1,2}\equiv |S_{1,2}|/m_{1,2}^2$.

\begin{figure}
\begin{center} 
\scalebox{0.45}{\includegraphics{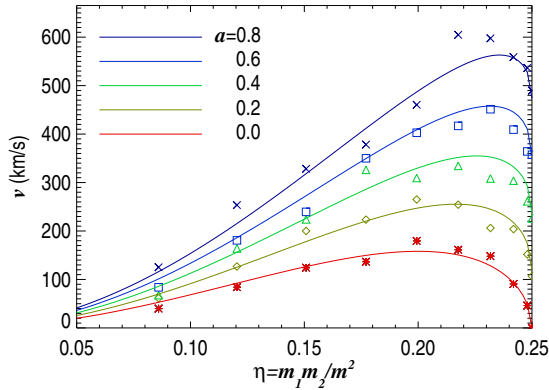}}\\
\caption{\label{plottwo} EOB predictions of recoil velocity for a range of
  unequal-mass BHs with spins aligned and anti-aligned with
  the orbital angular momentum. The solid lines are fits to equation
  (\ref{v_goddard}). To avoid confusion, we have not
  plotted error bars, which are typically $\sim 30-40\%$ in this case.}
\end{center}
\end{figure}

In Figure \ref{plottwo} we show the combination of the unequal-mass
and anti-aligned spin effects, with both black holes having equal
dimensionless spin magnitudes $a$. Following the 
approach of \citet{baker07}, we plot analytic fits to the EOB
predictions of the form 
\begin{equation}\label{v_goddard}
v(q,a) = \frac{32 V_0 q^2}{(1+q)^5}
\sqrt{(1-q)^2 + 2\,(1-q)\beta K_p + K_p^2 },
\end{equation}
where $q=m_2/m_1$ and $K_p=k_p a(q+1)$. We agree closely with their
best-fit results, matching $V_0=276$ km/s, and finding slightly
higher values of $\beta=1.27$, and $k_p=1.07$
(compare to their $\beta=0.84$ and $k_p=0.85$). This deviation is
largest for large spins, where they do not have much data 
and our simplified RD matching methods might start to break down.
Following the spin-orbit recoil formula in \citet{kidde95}, we can
modify equation (\ref{v_goddard}) to include non-planar kicks. 
We write the recoil component in the plane as
$K_p=k_p(a_1\cos\theta_1 - qa_2\cos\theta_2)$ and the component out
of the plane as $K_z=k_z(a_1\sin\theta_1\cos\phi_1 -
qa_2\sin\theta_2\cos\phi_2)$, with $\theta_{1,2}$ the angle between
the spin and angular momentum and $\phi_{1,2}$ the azimuthal angle of each
spin, measured with respect to the binary separation vector. From the
Monte Carlo simulations (described in the next Section), we find
$k_z\approx 1.7$, but with a large scatter between the EOB predictions and
the simple analytic model.

As a final check of our model, in Table \ref{tableone} we compare with
the more general 
configurations of \citet{koppi07}, \citet{campa07}, and
\citet{gonza07}. For the configuration of \citet{gonza07}, where
the spins are anti-aligned with each other and normal to the angular
momentum vector,
the value of the final recoil is sensitive to the angle the spins make with the
orbital velocity vector. We
maximize over this angle, but are still unable to attain their kick
magnitudes of $\sim 2500$ km/s. However, we see from these
different cases that there doesn't seem to be a {\it systematic}
disagreement with the NR results: sometimes the EOB
overestimates the kick,
and sometimes underestimates it, so the velocity
distributions integrated over a wide range of BH parameter space
should still be reasonably reliable.

\begin{table}
\caption{\label{tableone} Comparison of NR recoil calculations
with EOB model}
\vspace{0.1cm}
\begin{tabular}{cllccrrc}
\hline
\hline
$\frac{m_1}{m_2}$ & $a_1$ & $a_2$ & $\theta_1$ &$\theta_2$
& $v_{\rm NR}$ & $v_{\rm EOB}$ & Ref.\ \\
&  &  & (deg) & (deg) & (km/s) & (km/s) &  \\
\hline
1 & 0.58 & 0.0 & 0 & 180 & $100\pm 10$ & $120\pm 70$ & [1]\\
1 & 0.58 & 0.15 & 0 & 180 & $135\pm 20$ & $160\pm 100$ & [1]\\
1 & 0.58 & 0.29 & 0 & 180 & $185\pm 5$ & $260\pm 80$ & [1]\\
1 & 0.58 & 0.44 & 0 & 180 & $215\pm 15$ & $420\pm 210$ & [1]\\
\smallskip 
1 & 0.58 & 0.58 & 0 & 180 & $260\pm 15$ & $340\pm 160$ & [1]\\
\smallskip 
2 & 0.8 & 0.0 & 135 & 0 & $454\pm 25$ & $360\pm 150$ & [2] \\
1 & 0.73 & 0.73 & 90 & 90 & $2450\pm 250$ & $1700\pm 400$ & [3]\\
1 & 0.8 & 0.8 & 90 & 90 & $2650\pm 300$ & $1850\pm 450$ & [3]\\
\hline
\end{tabular} \\
{\bf References:} [1] \citet{koppi07}
[2] \citet{campa07}
[3] \citet{gonza07}
\end{table}

\section{RESULTS: RECOIL VELOCITY DISTRIBUTIONS}
\label{results}

Having successfully compared the EOB model with a range of NR results, we
are now in a position to carry out a series of Monte Carlo
calculations. The first model
we consider is that of equal-mass BHs with random spins,
uniformly distributed in $[0\le a \le 0.9]$ and $[-1 \le \cos\theta
\le 1]$.
We construct a probability distribution
function $P(v)$ for the kick velocities by summing over a collection of
normal distributions with the mean and variance for each
individual binary system calculated as in \S~\ref{analytic_models}.
The cumulative distribution function $f(v)$ is given by $f(v) =
\int_v^\infty P(v')dv'$, normalized such that $f(0)=1$ and
$f(v)$ is the probability that a random binary in the sample
will have a recoil larger than $v$. 

\begin{figure}
\begin{center} 
\scalebox{0.45}{\includegraphics{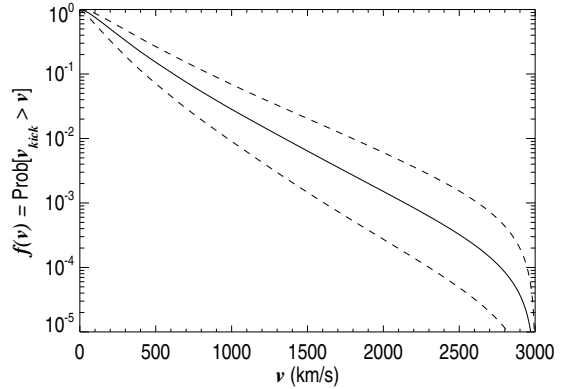}}\\
\caption{\label{plotthree} Cumulative distribution function for kick
velocities $v_{\rm kick} < v$ for equal masses $m_1=m_2$ and random
spins. The dashed lines represent 1-$\sigma$ confidence
limits.}
\end{center}
\end{figure}

Figure \ref{plotthree} shows this distribution for
the equal-mass, random spin case for $10^4$ binaries, with the dashed lines
corresponding to 1-$\sigma$ confidence limits. The
resulting uncertainty is on the order of $50\%$, which, while
admittedly large, can still provide interesting new constraints on
astrophysical models of black hole mergers. 

\begin{figure}
\begin{center} 
\scalebox{0.45}{\includegraphics{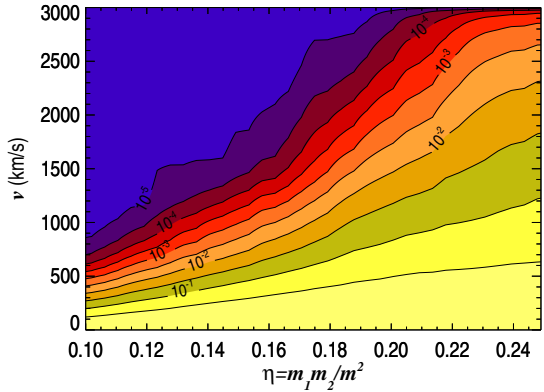}}\\
\caption{\label{plotfour} Cumulative probability distribution as a
function of symmetric mass ratio $\eta$. The contours of $f(v;\eta)$
represent the probability of having a recoil velocity greater than
$v$. The spins have amplitude $a_1=a_2=0.9$ and random orientation.}
\end{center}
\end{figure}

Perhaps a more realistic model samples a uniform distribution
in mass ratios $[1\le m_1/m_2 \le 10]$
and spin orientations, and set $a=0.9$ for each black hole, based on
observational arguments that most supermassive black holes are rapidly
spinning \citep{yu02,elvis02,wang06}. For these parameters, we find
a fraction $f_{\rm 500}=0.12^{+0.06}_{-0.05}$ of
binaries have recoil velocities
greater than 500 km/s and $f_{\rm 1000}=0.027^{+0.021}_{-0.014}$
have kicks greater than 1000 km/s. However, as we saw in
\S~\ref{compare_NR}, the largest kicks occur with nearly equal masses,
so these fractions may be biased towards
smaller kicks due to the wide range of $m_1/m_2$ considered in this
model. To factor out the uncertainty in mass distributions, we plot in Figure
\ref{plotfour} the velocity distribution function as a function of
$\eta$. As expected, the typical velocities
increas significantly for $\eta \gtrsim 0.16$. Limiting the sample
to this range of masses, we find $f_{\rm 500}=0.31_{-0.12}^{+0.13}$ and
$f_{\rm 1000}=0.079^{+0.062}_{-0.042}$.

\begin{figure} 
\begin{center}  
  \scalebox{0.45}{\includegraphics{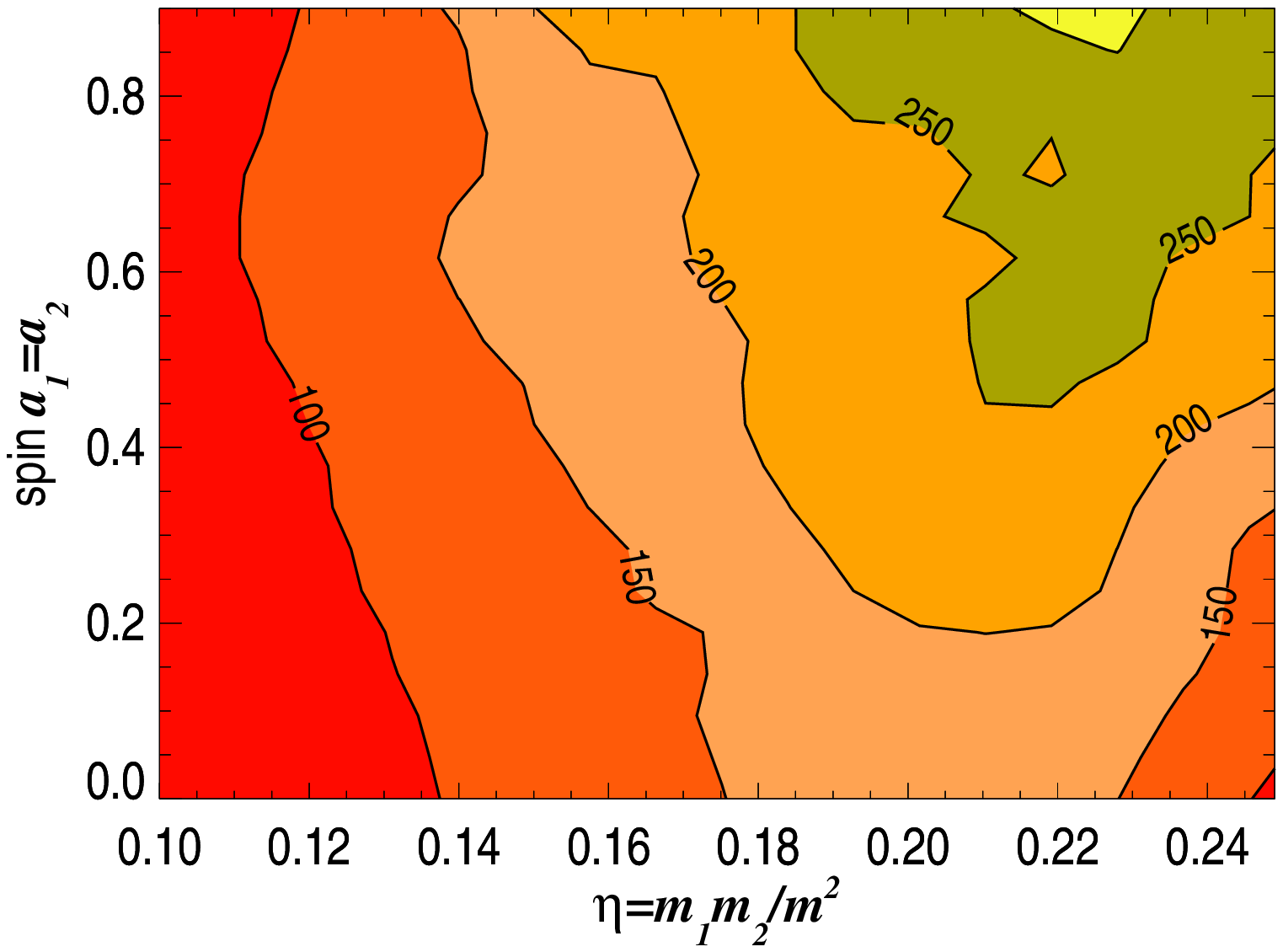}}\\
  \scalebox{0.45}{\includegraphics{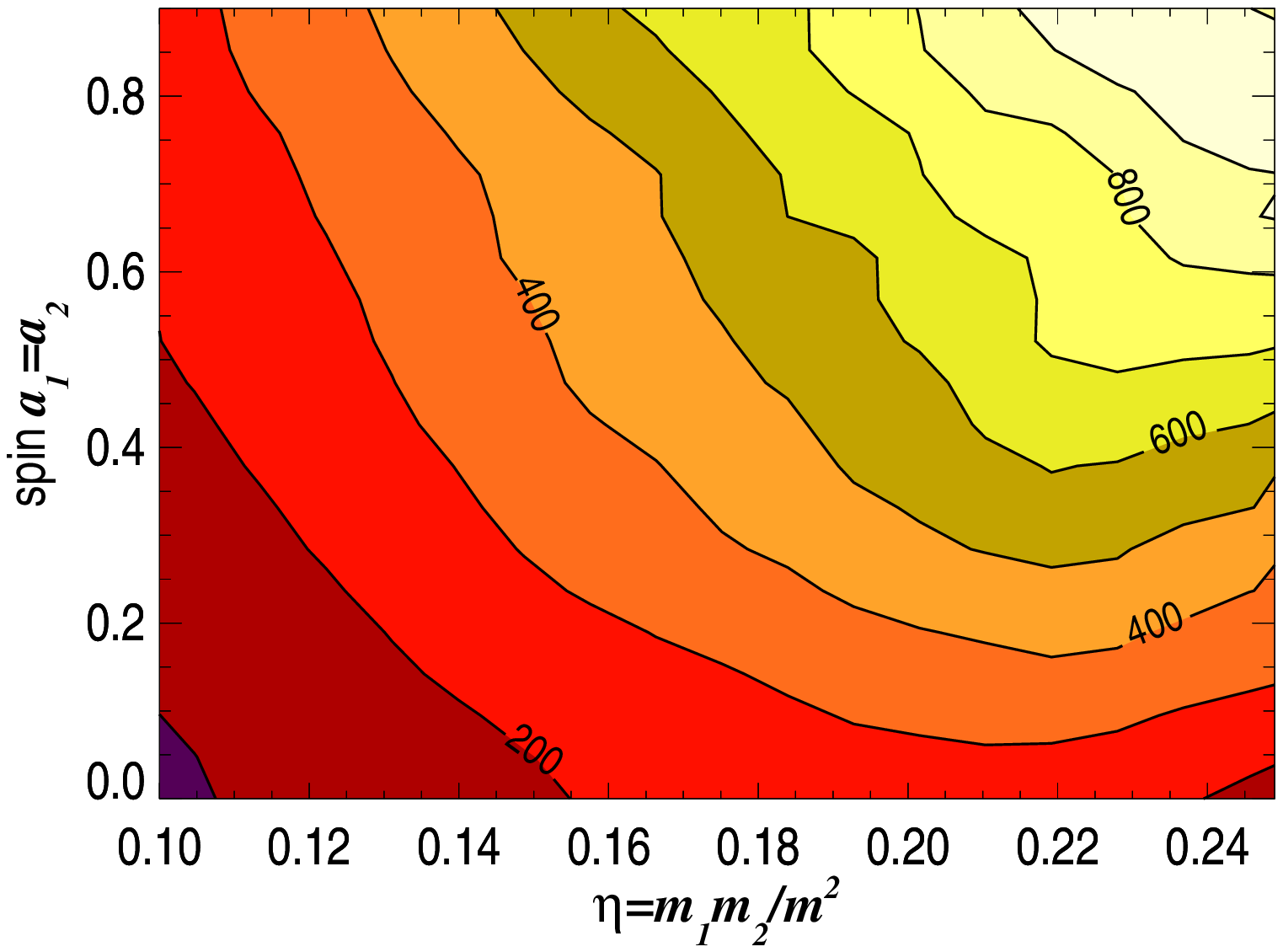}}
  \caption{\label{plotfive} \textit{Upper:} Contours of
    $v_{50}(\eta,a)$, defined 
such that $50 \%$ of BH binaries with a given mass ratio $\eta$ and spin
parameter $a$ are expected to have recoil velocities less than $v_{50}$. 
\textit{Lower:} Contours of $v_{90}(\eta,a)$, defined analogously to
$v_{50}$.}
\end{center}
\end{figure}

It is quite possible that black holes in the early universe have
actually gained most of their mass by mergers as opposed to accretion,
and thus may not be rapidly spinning \citep{hughe03}. To explore this
possibility, we performed another calculation
with the same range of mass ratios and equal spin
magnitudes $a_1=a_2$, uniformly distributed over $[0\le a_{1,2} \le 0.9]$, again with
random orientations. After calculating the resulting distribution
function $f(v;\eta,a)$, we define the functions $v_{50}(\eta,a)$ as
the velocity below which $50\%$ of the predicted kicks
lie. Similarly, $v_{90}(\eta,a)$ is greater than $90\%$ of the kick
velocities for a given $\eta$ and $a$. Contours of these
functions are shown in Figure \ref{plotfive}. As expected, we see that
systems with large spins and nearly equal masses have the highest
kicks.

We found by inspection that the distribution function $f(v;\eta,a)$
can be well approximated for $v \lesssim 1500$ km/s by 
\begin{equation}\label{f_v90}
f(v;\eta,a) \approx \exp\left[-\ln 10
  \left(\frac{v^2}{v_{90}^2(\eta,a)}\right)\right].
\end{equation}
Again motivated by maximizing the phenomenological equation
(\ref{v_goddard}), we found a relatively good analytic fit
\begin{equation}\label{v_90goddard}
v_{90}(q,a) \approx \frac{32 V_0 q^2}{(1+q)^5}
\sqrt{(1-q)^2 + k^2 a^2(q+1)^2},
\end{equation}
with $V_0 = 560$ km/s and $k = 1.3$, matching the EOB
prediction of $v_{90}$ within $25\%$ over the entire range of
$(\eta,a)$ investigated, and within $10\%$ for the great majority of
it. As a check of this method, we applied it to the other Monte Carlo
samples described above and along with equation (\ref{f_v90}) are able
to match closely the integrated distribution functions $f(v)$ for $v
\lesssim 1500$ km/s.

\section{DISCUSSION}
\label{discussion}

Along with the recent NR calculations of recoil velocities, there has
been a great deal of discussion of the astrophysical consequences of
these kicks. Here we only give a brief summary of these results. For a
more detailed review, see \citet{baker06,campa07}, and references
therein. 
Of particular interest to our calculation is the
observational evidence that most of the supermassive BHs
in the low-$z$ universe are rapidly spinning
\citep{yu02,elvis02,wang06}, and may 
have undergone multiple mergers with $m_1/m_2 \lesssim 3$
\citep{haehn02,merri06}, and thus have quite possibly received kicks
on the order of 500 or even 1000 km/s. This should be enough to eject
the final BH from spiral galaxy bulges and even most giant elliptical
galaxies \citep{merri04} [globular clusters may favor more extreme
mass ratios, but also have significantly smaller
escape velocities \citep{mille02}]. At the same time, there is some
observational evidence {\it against} such large kicks
\citep{libes06}. This apparent discrepancy may be explained with 
more detailed models for the BH mass distributions in hierarchical
merger scenarios, as well as the
possibility of using certain preferred spin orientations that would reduce
the kick magnitude \citep{schni04}. At the same time, to reduce the
uncertainties in our predictions, we plan to improve the EOB model,
notably the details of the matching method and the 
determination of the RD frequencies.

\vspace{0.25cm}\noindent A.B.\ acknowledges support from NSF grant
PHY-0603762 and from the Alfred Sloan Foundation.


\begin{thebibliography}{99}


\bibitem[Baker et al.(2006)]{baker06} Baker, J.\ G., Centrella, J.,
  Choi, D.-I., Koppitz, M., van Meter, J.\ R., \& Miller, M.\ C. 2006
  ApJ, 653, L93.

\bibitem[Baker et al.(2007)]{baker07} Baker, J.~G., Boggs, W.\ D.,
Centrella, J., Kelly, B.\ J., McWilliams, S.\ T., Miller, M.\ C., \&
van Meter, J.\ R. 2007, [astro-ph/0702390].

\bibitem[Blanchet et al.(2004)]{blanc04} Blanchet, L., Damour, T.,
  Esposito-Farese, G., \& Iyer, B.\ R. 2004, Phys.\ Rev.\ Lett.\ 93,
  091101.

\bibitem[Blanchet et al.(2005)]{blanc05} Blanchet, L., Qusailah, M.\ S.\ S., \&
Will, C.\ M. 2005, ApJ 635, 508.

\bibitem[Buonanno \& Damour(1999)]{BD1} Buonanno, A. \& Damour,
T. 1999, Phys.\ Rev.\ D 59, 084006.

\bibitem[Buonanno \& Damour(2000)]{BD2} Buonanno, A. \& Damour,
T. 2000, Phys.\ Rev.\ D 62, 064015.

\bibitem[Buonanno et al.(2006a)]{BCD} Buonanno, A., Chen, Y. \&
Damour, T. 2006, Phys.\ Rev.\ D 74, 104005.

\bibitem[Buonanno et al.(2006b)]{BCP} Buonanno, A., Cook, G. \&
Pretorius, T. 2006, [gr-qc/0610122].

\bibitem[Campanelli et al.(2007)]{campa07} Campanelli, M., Lousto,
C.O., Zlochower, Y. \&  Merritt, D. 2007, [gr-qc/0701164].

\bibitem[Damour et al.(2000)]{DJS} Damour, T., Jaranowski, P., \&
Sch\"afer, G. 2000, Phys.\ Rev.\ D 62, 084011. 

\bibitem[Damour \& Gopakumar(2006)]{damour06} Damour, T., \&
Gopakumar, A. 2006, Phys.\ Rev.\ D 73, 124006. 

\bibitem[Elvis et al.(2002)]{elvis02} Elvis, M., Risaliti, G., \&
  Zamorani, G. 2002, ApJ 565, L75.

\bibitem[Fitchett (1983)]{fitch83} Fitchett, M.\ J. 1983, MNRAS 203, 1049.

\bibitem[Flanagan \& Hughes(1998)]{flana98} Flanagan, E.\ E., \&
  Hughes, S.\ A. 1998, Phys.\ Rev.\ D 57, 4535.

\bibitem[Gonzalez et al.(2006)]{gonza06} Gonzalez, J.~A., Sperhake,
U., Br\"ugmann, B., Hannam, M.~D., \& Husa, S. 2006, [gr-qc/0610154]. 

\bibitem[Gonzalez et al.(2007)]{gonza07} Gonzalez, J.~A., Hannam, M.\ D.,
M.~D., Sperhake, U., Br\"ugmann, B., \& Husa, S. 2007, [gr-qc/0702052].

\bibitem[Haehnelt \& Kauffmann(2002)]{haehn02} Haehnelt, M.\ G., \&
Kauffmann, G. 2002, MNRAS 336, L61.

\bibitem[Herrmann et al.(2006)]{herrm06} Herrmann, F.,
Shoemaker, D., \& Laguna, P. 2006, [gr-qc/0601026]. 

\bibitem[Herrmann et al.(2007)]{herrm07} Herrmann, F., Hinder, I.,
Shoemaker, D., Laguna, P., \& Matzner, R.~A. 2007, [gr-qc/0701143]. 

\bibitem[Hughes \& Blandford(2003)]{hughe03} Hughes, S.\ A., \&
Blandford, R.\ D. 2003, ApJ 585, L101.

\bibitem[Kidder(1995)]{kidde95} Kidder, L.\ E. 1995, Phys.\ Rev.\ D
  52, 821.

\bibitem[Koppitz et al.(2007)]{koppi07} Koppitz, M., Pollney, D.,
Reisswig, C., Rezzolla, L., Thornburg, J., Diener, P. \& Schnetter,
E. 2007, [gr-qc/0701163].

\bibitem[Libeskind et al.(2006)]{libes06} Libeskind, N.\ I., Cole, S.,
  Frenk, C.\ S., \& Helly, J.\ C. 2006, MNRAS 368, 1381.

\bibitem[Merritt et al.(2004)]{merri04} Merritt, D., Milosavljevic,
  M., Favata, M., \& Hughes, S.\ A. 2004, ApJ 607, L9.

\bibitem[Merritt(2006)]{merri06} Merritt, D. 2006, ApJ 648, 976.

\bibitem[Miller \& Hamilton(2002)]{mille02} Miller, M.\ C., \&
Hamilton, D.\ P. 2002, MNRAS 330, 232.

\bibitem[Menou et al.(2001)]{menou01} Menou, K., Haiman, Z., \&
  Narayanan, V.\ K. 2001, ApJ 558, 535.

\bibitem[Sopuerta et al.(2006)]{sopue06} Sopuerta, C.\ F., Yunes, N., \&
Laguna, P. 2006, [astro-ph/0611110].

\bibitem[Schnittman (2004)]{schni04} Schnittman, J.\ D. 2004, Phys.\
Rev.\ D, 124020.

\bibitem[Thorne (1980)]{thorn80} Thorne, K.\ S. 1980, Rev.\ Mod.\ Phys.\
52, 299.

\bibitem[Wang et al.(2006)]{wang06} Wang, J.-M. 2006, ApJ 642, L111.

\bibitem[Yu \& Tremaine(2002)]{yu02} Yu, Q.\ J., \& Tremaine, S. 2002,
  MNRAS 335, 965.

\end{thebibliography}
\end{document}